\documentclass[%
 reprint,
 superscriptaddress,
 amsmath,amssymb,
 aps,
 prl,
]{revtex4-2}
\usepackage{graphicx}
\usepackage{dcolumn}
\usepackage{bm}
\usepackage{physics}
\usepackage{hyperref}
\usepackage[dvipsnames]{xcolor}
\usepackage[caption=false]{subfig}
\usepackage{svg}

\newif\ifdraft
\drafttrue

\ifdraft
\def\jrf#1{{\color{orange} [JRF: #1]}}
\def\sn#1{{\color{PineGreen} [SN: #1]}}
\def\ds#1{{\color{blue} [DS: #1]}}
\def\mc#1{{\color{purple} [MC: #1]}}
\else
\def\jrf#1{}
\def\sn#1{}
\def\ds#1{}
\def\mc#1{}
\fi

\renewcommand{\selectlanguage}[1]{}
\newcommand{\ketpsi}[1][]{\ket{\psi_{#1}}}

\newcommand{\ketphi}[1][]{\ket{\phi_{#1}}}
\newcommand{\braphi}[1][]{\bra{\phi_{#1}}}
\newcommand{\agp}{A_{\lambda}}
\newcommand{\exactagp}{\Tilde{A}_\lambda}
\newcommand{\dhdl}{\partial_{\lambda}H}
\newcommand{\vardhdl}{\mathrm{var}[\partial_{\lambda}H]_{\ketphi[0]}}
\newcommand{\omegamax}{\omega_{\max}}
\newcommand{\infnorm}[1]{\left\lVert#1\right\rVert_\infty}
\newcommand{\indexset}{\mathcal{I}_{\omegamax}}
\newcommand{\gap}{\Delta}
\newcommand{\deltaomega}{\delta_{\omega}}
\newcommand{\omegaavg}{\omega_{\mathrm{avg}}}

\newcommand{\sumgz}[1]{\sum_{#1>0}}

\begin{document}

\title{Counterdiabatic Driving with Performance Guarantees}

\author{Jernej Rudi Fin\v{z}gar}
\affiliation{BMW AG, Munich, Germany}
\affiliation{Technical University Munich, School of CIT, Department of Computer Science, Garching, Germany}
\email[Correspondence email address:\\]{jernej-rudi.finzgar@tum.de}

\author{Simone Notarnicola}
\affiliation{Department of Physics, Harvard University, Cambridge, Massachusetts 02138, USA}
\affiliation{Dipartimento di Fisica e Astronomia, Universit\`a degli Studi di Padova, I-35131 Italy}
\affiliation{Istituto Nazionale di Fisica Nucleare (INFN), Sezione di Padova, I-35131 Italy}

\author{Madelyn Cain}
\affiliation{Department of Physics, Harvard University, Cambridge, Massachusetts 02138, USA}

\author{Mikhail D. Lukin}
\affiliation{Department of Physics, Harvard University, Cambridge, Massachusetts 02138, USA}

\author{Dries Sels}
\affiliation{Center for Computational Quantum Physics, Flatiron Institute, 162 5th Avenue, New York, New York 10010, USA}
\affiliation{Center for Quantum Phenomena, Department of Physics, New York University, 726 Broadway, New York, New York 10003, USA}

\date{\today}

\begin{abstract}  
    Counterdiabatic (CD) driving has the potential to speed up adiabatic quantum state preparation by suppressing unwanted excitations. However, existing approaches either require intractable classical computations or are based on approximations which do not have performance guarantees. 
    We propose and analyze a non-variational, system-agnostic CD expansion method and analytically show that it converges exponentially quickly in the expansion order.
    In finite systems, the required resources scale inversely with the spectral gap, which we argue is asymptotically optimal. 
    To extend our method to the thermodynamic limit and suppress errors stemming from high-frequency transitions, we leverage finite-time adiabatic protocols.
    In particular, we show that a time determined  by the quantum speed limit is sufficient to prepare the desired ground state, without the need to optimize the adiabatic trajectory. Numerical tests of our method on the quantum Ising chain show that our method can outperform state-of-the-art variational CD approaches.
\end{abstract}

\maketitle

\emph{Introduction.}---Adiabatic transport is a fundamental tool for preparing quantum states~\cite{albash_adiabatic_2018}, with applications ranging from  quantum many-body  physics~\cite{Monroe_2021} to quantum computation~\cite{aharonov_adiabatic_2008}, and optimization~\cite{kadowaki_quantum_1998, farhi_quantum_2001, hauke2020perspectives}. In spite of many successful experimental realizations~\cite{albash2018demonstration, ad_superconducting_ETH_2019,  browaeys_ssh, ebadi_quantum_2021, ebadi_quantum_2022, king_coherent_2022, King2023, Manovitz2025}, its scalability to larger system sizes is limited by the required long adiabatic timescales associated with a vanishing energy gap~\cite{born_beweis_1928_corr}, which is at odds with the coherence times achievable in practical quantum simulation and computation platforms. 
These considerations have spurred significant interest in developing potential shortcuts to adiabaticity~\cite{guery-odelin_shortcuts_2019}.

One promising approach to shortcutting adiabaticity is counterdiabatic (CD) driving. Given a time-dependent Hamiltonian, in this approach one can add terms which explicitly counteract the non-adiabatic transitions generated by the time derivative of its eigenstates~\cite{demirplak_adiabatic_2003, berry_transitionless_2009, del_campo_shortcuts_2013, kolodrubetz_geometry_2017}. 
However, in general, these terms, which form the so-called Adiabatic Gauge Potential (AGP), are highly nonlocal and require full knowledge of the spectrum for exact implementation. This has motivated the development of \emph{variational CD driving}~\cite{sels_minimizing_2017}, where local approximations of the AGP are constructed~\cite{claeys_floquet-engineering_2019}. Since its inception, this method has been widely adopted for ground state preparation. It has been extensively analyzed and refined through numerical studies~\cite{passarelli_counterdiabatic_2020, prielinger_two-parameter_2021, xie_variational_2022, petiziol_quantum_2024, Barone2024, kadowaki_greedy_2023, passarelli_counterdiabatic_2023, cai_quantum_2024, ji_counterdiabatic_2022, hartmann_polynomial_2022, gangopadhay_counterdiabatic_2024, schindler_counterdiabatic_2024, gjonbalaj, Hatomura2021, cepaite_counterdiabatic_2023} and successfully demonstrated in experimental quantum platforms~\cite{zhang_analog_2024, zhou_experimental_2020, kumar_digital-analog_2025}. However, the resource requirements and convergence guarantees associated with such approaches remain poorly understood, despite recent progress in digital implementations of CD driving~\cite{vanVreumingen2024}.

This Letter introduces a new method of constructing CD drives with rigorous performance guarantees. 
We derive a system-agnostic scheme to approximate the exact AGP, which makes the resource estimation and the error bounds universal.
In our approach, the resources scale inversely with the spectral gap between the ground and the first excited state $\gap$, and at worst linearly in the largest excitation frequency in the system. We argue that in generic, nonintegrable systems, such scaling is optimal up to subleading corrections. 
Because of the scaling with the largest excitation frequency, the performance guarantees are inapplicable in the thermodynamic (TD) limit, even if the system remains gapped. 
We show that this can be remedied by adiabatically suppressing high-frequency transitions by scaling the protocol’s duration with $1/\gap$. 
This approach allows one to effectively reach the quantum speed limit~\cite{roland_quantum_2002} without the need for optimizing the adiabatic trajectory. 
Our framework allows one to systematically address the challenges associated with adiabatic preparation of many body states and reduces the associated errors by increasing nonlocality. 

\begin{figure}[tb]
    \centering
    \includegraphics{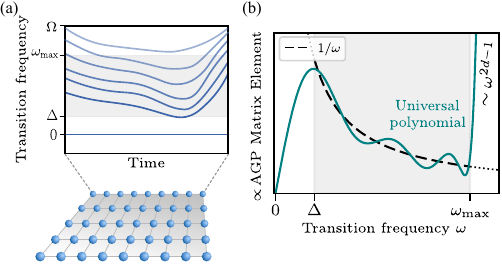}
    \caption{
    \textit{Universal counterdiabatic driving.}
    (a) 
    Our CD scheme seeks to suppress transitions from the ground state to the rest of the eigenspectrum during adiabatic quantum many-body state preparation.
    The figure shows the transition frequencies as a function of time with $\Delta$ the minimum spectral gap and $\Omega$ the largest transition frequency. In our scheme, we target the transitions with frequencies in the interval $[\Delta, \omegamax]$ (grey shaded area), where $\omegamax$ is a cutoff frequency chosen to enhance the protocol's efficiency.
    (b) Universal CD driving suppresses these transitions by approximating the  adiabatic gauge potential (AGP) matrix elements over these transition frequencies by a polynomial of degree $2d-1$, whose coefficients depend only on the interval bounds $[\gap,\omegamax]$. For frequencies $\omega>\omegamax$, the polynomial grows as $\omega^{2d-1}$.
    }
    \label{fig:poly-fit-one-over-x}
\end{figure}

\emph{Finite systems.}---
Our goal is to prepare the ground state of a given Hamiltonian $H$ by means of an adiabatic protocol with duration $\tau$ controlled by the parameter $\lambda(t)$ (see Fig.~\ref{fig:poly-fit-one-over-x}(a)). The initial Hamiltonian and final Hamiltonian correspond to $\lambda(t=0)=0$ and $\lambda(t=\tau)=1$, respectively. Let $\{\ketphi[n]\}$ denote its instantaneous eigenbasis.
A counterdiabatic protocol consists of evolving the ground state at $t=0$ under the equation
\begin{equation}
\label{eq:schrodinger}
    i\partial_t\ketpsi 
    =
    \left[
    H(\lambda)
    +
    \dot{\lambda}(t) \agp
    \right]
    \ket{\psi},
\end{equation}
where $\agp$ is the chosen counterdiabatic potential. If $\agp = \exactagp$, where $\exactagp$ is the exact AGP with matrix elements
\begin{equation}
\label{eq:AGP_exact}
    \mel{\phi_m}{\exactagp}{\phi_n}
    =
    i\mel{\phi_m}{\partial_\lambda}{\phi_n}
    =
    -i\frac{\mel{\phi_m}{\dhdl}{\phi_n}}{\omega_{mn}},
\end{equation}
with the transition frequencies $\omega_{mn}:=E_m-E_n$, the system will remain in the ground state at all times $t$, because the exact AGP precisely cancels out the transitions from the ground state~\cite{berry_transitionless_2009}.

Our goal is to suppress diabatic transitions by approximating the matrix elements of the exact AGP $\exactagp$~\eqref{eq:AGP_exact}. We parametrize the approximate AGP $\agp$ as
\begin{equation}
\label{eq:AGP_approx}
\mel{\phi_m}{\agp}{\phi_n}
=
-i\,p(\omega_{mn})
\mel{\phi_m}{\dhdl}{\phi_n},
\end{equation}
where $p(\omega_{mn})=\sum_{k=1}^d c_k \omega_{mn}^{2k-1}$ is a polynomial in the transition frequency.
To realize a practical implementation of Eq.~\eqref{eq:AGP_approx}, we leverage the nested commutator expansion~\cite{claeys_floquet-engineering_2019}
\begin{equation}
\label{eq:nested_commutator_expansion}
    \agp
    =
    i
    \sum_{k=1}^d
    c_k 
    \underbrace{
    [
    H,
    [
    H,\dots
    [
    H,}_{2k-1} \dhdl
    ]
    ]
    ],
\end{equation}
where each nested commutator of order $2k-1$ provides the requisite matrix elements ${\propto \mel{\phi_m}{\dhdl}{\phi_n}\omega_{mn}^{2k-1}}$ of $\agp$~\eqref{eq:AGP_approx}. Consequently, the remaining task is to find coefficients $c_k$ that best approximate the exact AGP.

Previously, the coefficients $c_k$ were determined by variational minimization~\cite{sels_minimizing_2017} (see also SM~\cite{supp}). In contrast, the central idea of our scheme is to analytically construct a system-agnostic, non-variational polynomial $p(\omega_{mn})$ in Eq.~\ref{eq:AGP_approx} to approximate the $1/\omega_{mn}$ dependence of the exact AGP matrix elements over a chosen range of relevant transition frequencies (see Fig.~\ref{fig:poly-fit-one-over-x}). 
The only input required to determine the polynomial coefficients will be the relevant range of frequencies $[\gap, \omegamax]$, where $\gap$ is the minimum spectral gap and $\omegamax$ is a chosen cut-off frequency. Although $\gap$ might not be precisely known in practice, it suffices to proceed with (conservative) estimates, e.g., based on known properties of the system. The effect of potentially overestimating $\gap$ depends on the problem-specific structure of low-frequency transitions. However, because the polynomial specifying $\agp$ decays to zero for $\omega<\gap$ (see Fig.~\ref{fig:poly-fit-one-over-x}(b)) a suboptimal choice is not expected to incur additional errors compared to a bare adiabatic evolution.

In summary, our construction does not require detailed information about the system Hamiltonian, and allows us to link the resource requirements of CD driving to the spectral gap $\Delta$.

We have thus reformulated the task of constructing an approximate AGP as a problem of polynomial approximation. Chebyshev polynomials are the canonical choice for approximating functions on bounded intervals, due to their nearly optimal convergence properties~\cite{achiezer_theory_1956}. To ensure that $\agp$ is Hermitian, we require that $p(\omega_{mn})$ be odd (see Eq.~\eqref{eq:AGP_approx}). We adopt the method of Hasson \& Restrepo~\cite{hasson_approximation_2007}, which employs a Chebyshev expansion to construct a near-optimal odd polynomial approximation of $1/\omega_{mn}$ on a given interval (see Appendix for details). Crucially, the coefficients depend solely on the bounds of the approximation interval, underscoring the system-agnostic nature of the approach.

\begin{figure*}
    \centering
    \includegraphics[]{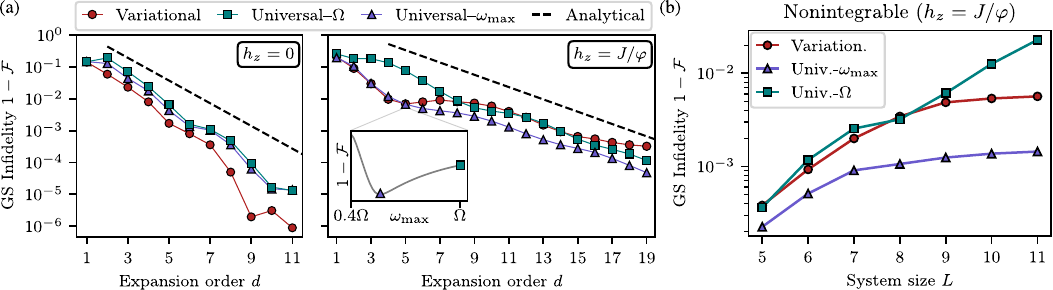}
    \caption{\textit{Universal CD driving at finite sizes.} (a) Ground-state infidelity for an integrable and nonintegrable point of the Ising model with $L=6$ in the limit of $\tau\to 0$, as a function of the expansion order $d$. The universal scheme is displayed for the choice $\omegamax=\Omega$ (teal squares) and for the optimal choice of $\omegamax$ at each expansion order $d$ (see inset in (a) for $d=5$ and purple triangles).
    Our scheme has comparable performance to variational AGP (red circles). The analytical scaling from Eq.~\eqref{eq:infidelity-only-poly} is shown in dashed black.
    (b) System size scaling of the infidelity in the nonintegrable case for $d=14$.}
    \label{fig:numerics-tau0}
\end{figure*}

The error of the polynomial approximation at $d$th order is directly related to the ground state fidelity at the end of the protocol.  By setting $\omegamax=\Omega$, where $\Omega$ is the largest transition frequency, we expand $1/\omega_{mn}$ over all excitations in the system and arrive at the following upper bound on the ground state infidelity (see Appendix):
\begin{equation}
\label{eq:infidelity-only-poly}
    1-\mathcal{F}
    \leq
    C\,
    (\log d)^2
    \left(
    \frac{
    1 - \gap/\Omega
    }{
    1 + \gap/\Omega
    }
    \right)^{2(d-1)}
    \hspace{-0.1cm}
    \vardhdl,
\end{equation}
with $\vardhdl = \braphi[0](\dhdl)^2\ketphi[0]-\braphi[0](\dhdl)\ketphi[0]^2$. Expanding the term $(\cdot)^{2(d-1)} =: x = \exp[\log x]$, we find that for $\Delta\ll\Omega$ 
the ground state infidelity scales proportionally to 

\begin{equation}
\label{eq:infidelity-only-exp}
1-\mathcal{F} \sim \exp[-4\gap d/\Omega],
\end{equation}
which means that the required degree asymptotically scales as 
$
    d\sim \Omega/\gap,
$
where we have omitted the logarithmic contribution from $d$, as well as from $\vardhdl$. 
In most cases, the $1/\gap$ dependence will be the dominant contribution. Remarkably, this dependence on the gap is similar to that found for the required duration of optimized adiabatic protocols~\cite{roland_quantum_2002}, where the schedule function $\lambda$ is chosen according to the gap profile. Here we obtain a similar scaling without the need for such optimization.

To showcase our scheme, we consider the one-dimensional transverse field Ising chain 
\begin{equation}
\label{eq:ising}
H(\lambda)
=
J\sum_{j=1}^{L-1}\sigma_i^{z}\sigma_{i+1}^{z}
+
h_z \sum_{j=1}^{L}\sigma_i^{z}
+
h_x(\lambda) \sum_{j=1}^{L}\sigma_i^{x},
\end{equation}
with a linear ramp connecting ${h_x(0) = J/2}$ and ${h_x(1) = 5J/2}$.
For simplicity, we choose the limit $\tau\to 0$, since Equation~\eqref{eq:infidelity-only-poly} is valid for any protocol duration $\tau$. In this limit, the original Hamiltonian drops out, leaving evolution governed only by the approximate AGP, so its influence can be examined independently of finite-time adiabatic effects. We compute both the variational AGP~\cite{sels_minimizing_2017, claeys_floquet-engineering_2019, supp} and our universal AGP at different orders $d$, and use them to evolve the ground state at $\lambda=0$ under Eq.~\eqref{eq:schrodinger}. 
In Fig.~\ref{fig:numerics-tau0}(a), we display the ground state infidelity for two scenarios: the integrable point ($h_z=0$, within the positive parity subspace), and for the generic nonintegrable regime ($h_z=J/\varphi$, with $\varphi$ the golden ratio). The variational results are represented by red circles, while the universal results with $\omegamax = \Omega$ are shown as teal squares. 
In purple we show results obtained using $\omegamax < \Omega$, which we discuss below.

As shown in Fig.~\ref{fig:numerics-tau0}(a), the infidelity of the universal scheme decreases exponentially with the expansion order $d$, in line with the analytic scaling predicted by  Eq.~\eqref{eq:infidelity-only-poly} (black dashed lines). 
Overall, the universal scheme achieves comparable performance to the variational method, despite being independent of the details of the Hamiltonian.
Both schemes exhibit a faster convergence for the integrable case, which can be attributed to the fact that it has fewer transitions and a smaller $\Omega$ (see Fig.~\ref{sfig:int-vs-nonint-structure}).
The variational scheme also exhibits a small relative advantage in this case, as it exploits the structure in the transition frequencies~\cite{claeys_floquet-engineering_2019, supp}, whereas the universal approach chooses the best \textit{uniform} approximation of the matrix elements of the exact AGP. 
In the non-integrable case, the schemes have comparable performance, as the 
transitions from the ground state are uniformely distributed in the $[\Delta, \Omega]$ interval.
Following this reasoning, it seems unlikely that much is to be gained from variational minimization for generic systems with chaotically distributed spectra.

\begin{figure*}[htb]
    \centering
    \includegraphics{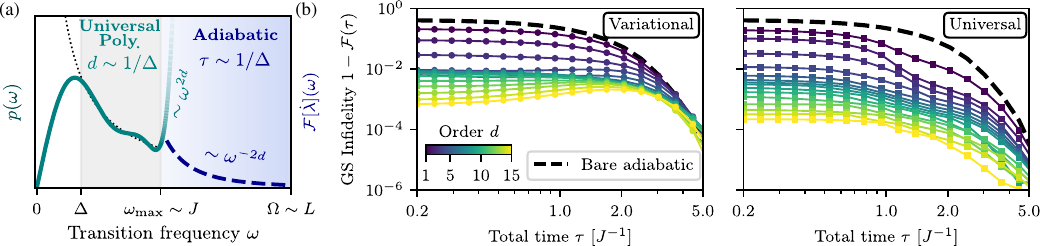}
    \caption{\textit{Universal CD driving at finite timescales.} (a) Combining the universal polynomial approximation of order $d\sim 1/\gap$ with adiabatic protocols of duration $\tau\sim 1/\gap$ allows us to construct a scheme that converges in the TD limit. The cut-off frequency $\omegamax$ can be chosen as some system size independent local energy scale $J$. Above $\omegamax$ the transitions are compensated by the frequency space decay of $\dot{\lambda}$, here shown in purple.
    (b) We show the time fidelities for the Universal and Variational approach for the nonintegrable Ising model with $h_z=J/\varphi$ and $L=6$. Fidelities of a bare adiabatic evolution plotted (dashed black lines) are plotted for comparison. For large $\tau$ the universal scheme improves upon the variational approach due to the additional flexibility of tuning $\omegamax$ as $\tau$ is increased. 
    }
    \label{fig:finite-times}
\end{figure*}

According to Eq.~\eqref{eq:infidelity-only-poly}, the quality of the polynomial approximation decreases with the ratio $\gap/\Omega$.
To observe this behavior, we fix the order $d$ and increase the system size, as this results in a smaller gap $\gap$ and a larger $\Omega$. The results in Fig.~\ref{fig:numerics-tau0}(b) confirm the exponential decrease in performance with increasing $\Omega/\gap$ (teal squares), as predicted by Eq.~\eqref{eq:infidelity-only-poly}. 

To improve practical performance, we therefore consider lower cut-off frequencies $\omegamax<\Omega$. Such a choice has two competing effects: on the one hand, it enhances the quality of the approximation in the interval $[\gap,\omegamax]$ for a given order $d$. 
On the other hand, it introduces a residual error $\varepsilon_{\text{res}}$ originating from the high frequency, ultraviolet (UV) transitions at $\omega>\omegamax$. At these frequencies the polynomial grows like $p(\omega) \sim \omega^{2d-1}$ and becomes a poor approximation of $1/\omega$ (see Fig.~\ref{fig:poly-fit-one-over-x}(b)). 
This implies the existence of an optimal $\omegamax$ that minimizes the infidelity at each $d$ (see inset in Fig.~\ref{fig:numerics-tau0}(a)). 
In Fig.~\ref{fig:numerics-tau0} we numerically show that the performance can be significantly improved with this choice of $\omegamax$.
Due to the improvements we observe by choosing $\omegamax<\Omega$, in the following we study how this choice affects the error bounds.

\emph{Convergence in the TD limit.}---
We have shown numerically that choosing $\omegamax<\Omega$ can lead to significant performance gains. 
Furthermore, the choice $\omegamax = \Omega$ is suboptimal in the TD limit where $\Omega = \mathcal{O}(L)$, as one would need $d\sim\mathcal{O}(L)$ even for gapped systems. 
This motivates us to explore the possibility of choosing $\omegamax \sim\mathcal{O}(1) < \Omega$.
This choice requires placing a bound on the residual error for frequencies above $\omegamax$. One might hope that such a bound can be achieved by noting that for generic systems the matrix elements for high frequency transitions decay exponentially~\cite{avdoshkin_euclidean_2020}. 
Unfortunately, exponential decay is not fast enough to control the error (see SM~\cite{supp} and Ref.~\cite{morawetz25}). 
To see this, note that for generic systems, the matrix elements of local operators such as $\braphi[n]\dhdl\ketphi[0]$, which govern the high frequency behavior of the AGP (see Eq.~\eqref{eq:AGP_approx}), decay (at least) exponentially as $\sim e^{-\gamma\omega_{n0}}$~\cite{avdoshkin_euclidean_2020}. 
In SM~\cite{supp}, we show that the corresponding residual error $\varepsilon_{\text{res}}$ scales as \mbox{$\sim \exp(-\gamma\omegamax)\omegamax^{2d}$}. 
Combining this result with the condition $d\sim \omegamax/\gap$ from Eq.~\eqref{eq:infidelity-only-exp}, we get $\varepsilon_{\text{res}}\sim d^d \exp(-d)$, i.e., our bound on the residual error scales factorially at large values of $d$.
While for some integrable models the decay is stronger and the factorial error can be suppressed (see SM and Ref.~\cite{morawetz25} for a detailed discussion), in generic systems we need further constraints to get favorable bounds for $\varepsilon_{\text{res}}$.

We will deal with the residual error by combining our scheme with a finite adiabatic timescale. More precisely, we will choose $\omegamax < \Omega$ and adiabatically suppress the transitions above $\omegamax$. The resulting residual error from transitions with frequencies $\omega>\omegamax$ (cf. Appendix Eq.~\eqref{eq:infidelity-general}) is proportional to the growth of the approximating polynomial $\sim \omega^{2d}$ at $\omega>\omegamax$ (see Fig.~\ref{fig:finite-times}(a)).
In SM~\cite{supp}, we show by means of adiabatic perturbation theory~\cite{chandra_adiabatic_2010} that by choosing sufficiently smooth and slow protocols, this growth can be compensated adiabatically. 
In particular, we use the smoothness of the protocol $\lambda(t)$, and its corresponding decay properties in frequency space, to offset the error stemming from the growth of the polynomial (see Fig.~\ref{fig:finite-times}(a)).
We find that such protocols require a total time $\tau \sim d$. Thus, by setting the cutoff frequency as some system-size-independent local energy scale $\omegamax\sim J$, and recalling the bound on approximation error from Eq.~\eqref{eq:infidelity-only-poly}, we find that 
\begin{equation}
        d\sim 1/\gap
        \quad\text{\&}\quad
        \tau\sim 1/\gap
\end{equation}
are required for the scheme to converge. Remarkably, this implies that one can achieve the quantum speed limit $\tau\sim 1/\gap$ without optimizing the adiabatic trajectory, provided that the necessary $d\sim 1/\gap$ CD terms can be engineered.

In Fig.~\ref{fig:finite-times}(b), we show the effects of adiabaticity for the transverse field Ising chain Eq.~\eqref{eq:ising}, with ${\lambda(t)=\sin^2\left(\frac{\pi}{2} \sin^2\left(\frac{\pi t}{2 \tau}\right)\right)}$. We plot the ground state infidelity for different expansion orders $d$ with finite protocol times $\tau$. For variational CD driving (left), the fidelities coalesce with the bare adiabatic fidelities at larger values of $\tau$,  thus limiting the space for improvement. In contrast, the universal CD driving (right) benefits from the additional flexibility of varying $\omegamax$, which can be chosen to optimize the performance for a given order $d$ and protocol duration $\tau$. This allows us to significantly improve upon the bare adiabatic protocol even for longer, nearly adiabatic, time scales.

Alternatively, one can still find universal schemes that remain well-defined in the TD limit, without invoking adiabaticity, by considering appropriate expansions of the exact AGP $\exactagp$. 
We begin by showing in the Appendix that the exact AGP $\exactagp$ can be expanded in a Krylov basis of the Liouvillian $\mathcal{L}=[H,\cdot]$ applied to $\dhdl$, i.e. a Gram-Schmidt orthogonalized version of the nested commutators in Eq.~\eqref{eq:nested_commutator_expansion}~\cite{takahashi_shortcuts_2024}. 
Using the universal properties of the high frequency tails of the spectral function~\cite{avdoshkin_euclidean_2020}, which translate to universal dynamics on the Krylov chain~\cite{parker19}, we show that one needs to keep least the lowest $d\sim e^{1/\gap}$ nested commutators in order for the error to remain small.

The complexity of exactly computing the Krylov expansion grows exponentially in $d$, so it becomes intractable rather quickly. 
As a solution, instead of performing the exact expansion, we exploit the exponential decay of matrix elements~\cite{avdoshkin_euclidean_2020} to build a new family of orthogonal polynomials.
We choose these polyonimals --- which we dub \emph{Laplace polynomials} --- to be orthogonal with respect to an exponentially decaying weight function. 
This allows us to define them on the entire real line, thus eliminating the residual error and the aforementioned factorial growth problem from the start.

In the SM~\cite{supp}, we construct the Laplace polynomials numerically, and express $1/\omega$ in them. We numerically confirm that this scheme exhibits the same resource requirements in the degree $d$ as the Krylov construction. Without any particular structure in the spectral function or Lanczos coefficients, this scaling cannot be improved.

\emph{Discussion \& Outlook.---}
In this work, we constructed a universal approach to CD driving based on generic information about the system, with rigorous error bounds and resource requirements. 
The resource cost of our approach depends on  the minimum spectral gap and the high-frequency transitions from the ground state to the rest of the spectrum.
We find that the effect of these transitions can be mitigated by leveraging adiabaticity. 
The resulting protocol reaches the quantum speed limit without the need for optimization, thus providing a practical recipe for improving performance which is robust in the TD limit. 
We numerically observe that this approach also outperforms state-of-the-art variational CD in the adiabatic limit. 
The demanding scaling of the required long-range terms in our constructions can pose a significant challenge for realistic implementations. However, if such nonlocality can be engineered, our framework can be used to exponentially reduce the error.

The present work can be extended in several directions. 
Our analytic bounds show that CD driving performance is inextricably linked to nonlocality.
While such nonlocality may be challenging to engineer experimentally, our work underscores the importance of identifying implementation strategies where CD driving gives a practical advantage.
To this end, Floquet engineering the nested commutators forming the Ansatz in Eq.~\eqref{eq:nested_commutator_expansion} may provide a route towards experimental implementation of approximate AGPs in quantum simulators~\cite{claeys_floquet-engineering_2019, köylüoğlu2024floquetengineeringinteractionsentanglement, visuri2025digitizedcounterdiabaticquantumcritical, Kiefer2019, liang2024floquetengineeringanisotropictransverse}.
In the digital setting, these terms could  be incorporated in Trotterized circuits using multiqubit controlled phase gates~\cite{kalinowski2023, Evered2023, Maskara2025,PRXQuantum.5.020362}. 
Furthermore, it would be interesting to design hybrid protocols combining CD driving with methods to circumvent small spectral gaps, such as exploiting diabatic transitions~\cite{Bernien2017,Crosson_2021, King2023}, using quantum quench algorithms~\cite{Hastings_2019, Schiffer2024, lukin2024quantumquenchdynamicsshortcut}, or exploring more favorable adiabatic paths~\cite{farhi2010quantumadiabaticalgorithmssmall, dickson_amin, lanting_king, cain_quantum_2023, finzgar_designing_2024}.
Identifying how to optimally implement such strategies while maintaining hardware efficiency can therefore enable a broad range of new practical applications of CD protocols.

\emph{Acknowledgements.---}
J.R.F. \& S.N. would like to thank Nik Gjonbalaj, Luka Pavešić \& Rahul Sahay for insightful discussions. J.R.F. \& S.N. would like to thank Oliver Lunt for useful comments on the derivation of the error bounds.
Numerical simulations in this work were performed using the \texttt{QuTiP} library~\cite{johansson_qutip_2012, johansson_qutip_2013, lambert_qutip_2024}. S.N. acknowledges funding from the European Union’s Horizon Europe research and innovation program under the Marie Sklodowska-Curie Grant No. 101059826 (ETNA4Ryd) and support from the Quantum Computing and Simulation Center (QCSC) of Padova University. 
M.C., S.N, and M.L. acnowledge support from the Department of Energy QSA (grant number DE-AC02-05CH11231) and QUACQO (grant number DE-SC0025572), National Science Foundation (grant numbers PHY-2012023 and CCF-2313084) and the Center for Ultracold Atoms (an NSF Physics Frontiers Center).
D.S. is grateful for ongoing support through the Flatiron Institute, a division of the Simons Foundation, and to AFOSR for support through Award no. FA9550-25-1-0067.

%

\begin{appendix}

\section{Details of the polynomial approximation}
\label{app:poly-approx}
The following draws heavily on the technique introduced in Ref.~\cite{hasson_approximation_2007}. The procedure is as follows; first, we will approximate $1/\omega$ on the interval $[\gap,\omegamax]$ without observing the odd parity requirement. We will then use the obtained series expansion and augment it to arrive at an odd polynomial expansion of the inverse function that will consequently work on $[-\omegamax, -\gap]\cup [\gap,\omegamax]$. It can be shown that the resulting approximation is optimal up to logarithmic factors.

\label{app:noparity-approx}
For convenience, we first map the domain of the function to $[-1,1]$ by defining $x=(\omega-\omegaavg)/\deltaomega$ where ${\omegaavg=(\omegamax+\gap)/}2$ and $\deltaomega=(\omegamax-\gap)/2$. Then, the goal is to find a near-optimal (in the $\infnorm{\cdot}$ sense) polynomial approximation of the function ${f(x)=1/(\deltaomega\cdot x + \omegaavg)}$ for $x\in [-1,1]$.

Chebyshev polynomials $T_n$, defined by the relation $T_n(\cos\theta)=\cos(n\theta)$, constitute an orthonormal basis with respect to the inner product ${\langle f, g\rangle = \int_{-1}^{1}\dd xf(x)g(x)(1-x^2)^{-1/2}}$. In particular, in Ref.~\cite{mathar_chebyshev_2006} series expansions of inverse polynomials are computed. We are interested in the series expansion of ${f(x)=1/(\deltaomega\cdot x + \omegaavg)}$. Defining $\eta:=\omegaavg/\deltaomega$ it can be shown that
\begin{equation}
\label{eq:series_expansion}
    f(x)
    =
    \frac{1}
    {\deltaomega\cdot(x+\eta)}
    = 
    \sum_{j=0}^{\infty} c_j T_j(x),
\end{equation}
where
\begin{equation}
\label{eq:series_coefs}
    c_j=\begin{cases}
    \frac{-1}{\deltaomega\sqrt{\eta^2 - 1}},
    &\text{for } j=0
    \\
    \frac{-2}{\deltaomega\sqrt{\eta^2 - 1}}\left(\frac{-1}{\eta+\sqrt{\eta^2 - 1}}\right)^j,
    &\text{else.} 
\end{cases}
\end{equation}

It can be shown~\cite{hasson_approximation_2007} that truncations of this infinite series expansion at $d$-th order satisfy
$$
    \infnorm{1/\omega- Q_d(\omega)}
    \leq
    C^\prime
    \left(
    \frac{
    1 - \sqrt{\frac{\gap}{\omegamax}}
    }{
    1 + \sqrt{\frac{\gap}{\omegamax}}
    }
    \right)^d,
    \vspace{0cm}
$$
where we have used $Q_d(\omega) = \sum_{j=0}^{d}c_j T_j(\deltaomega\omega + \omegaavg)$ and $C^{\prime}\in\mathbb{R}$ is some real constant. Moreover, it has been shown~\cite{hasson_approximation_2007} that this error is optimal up to a logarithmic factor, i.e., 
$$
    \infnorm{1/\omega- Q_d(\omega)}
    = 
    \inf_{p\in\mathcal{P}_d}
    \infnorm{1/\omega- p(\omega)} 
    \log{d}
$$
where $\mathcal{P}_d$ is the set of all real polynomials of degree $d$. 

We now use the series expansion above to obtain an \emph{odd} approximating polynomial. Following Ref.~\cite{hasson_approximation_2007} we perform the following steps:
\begin{enumerate}
    \item We approximate $1/\omega$ on $[\gap^2, \omegamax^2]$ by $Q_{d-1}(\omega)$.
    \item Then, $Q_{d-1}(\omega^2)$ approximates $1/\omega^2$ on $[\gap, \omegamax]$.
    \item Finally, $P_{2d-1}(\omega):=\omega\cdot Q_{d-1}(\omega^2)$ approximates $1/\omega$ on $[\gap, \omegamax]$. Here $d\geq 1$.
\end{enumerate}
We note that $\omega Q_{d-1}(\omega^2)$ is an odd polynomial (of degree $2d-1$) as it is a product of the odd polynomial $\omega$ and the even polynomial $Q_{d-1}(\omega^2)$. In Ref.~\cite{hasson_approximation_2007} it is shown that the error of such an approximation scales as
$$
    \infnorm{P_{2d-1}(\omega) - 1/\omega}
    \leq
    C \left(
    \frac{1 - \gap/\omegamax}{1 + \gap/\omegamax}
    \right)^{d-1} \log {d}.
$$
Moreover, it is shown that this is the best possible polynomial approximation up to the $\log d$ factor.

\section{Decomposing the GS infidelity}

We can rewrite the Schrödinger Eq.~\eqref{eq:schrodinger} in the instantaneous eigenbasis of $H(\lambda)$ using the ansatz ${\ketpsi=\sum_n a_n(t) \ket{\phi_n(t)}}$ as
\begin{equation}
\label{eq:schrodinger-eigenbasis}
    i\partial_t a_n 
    - 
    \dot{\lambda}
    \sum_{m} a_m 
     \braphi[n](\exactagp-\agp)\ketphi[m]
    =
    E_n a_n,
\end{equation}
where we have used the definition of the exact AGP $i\braphi[n]\partial_t\ketphi[m]=\dot{\lambda}\braphi[n]\exactagp\ketphi[m]$.

By employing the gauge transformation ${\alpha_n=a_n\exp\left[i\Theta_n(t)\right]}$ where $\Theta_n(t)=\int_0^t\dd t' E_n(t')$ we can obtain the following integral equations for eigenstate populations at the end of the protocol~\cite{chandra_adiabatic_2010}
\begin{equation}
    \label{eq:int-schrodinger}
    \alpha_n(\tau)
    \hspace{-0.03cm}
    =
    \hspace{-0.03cm}
    \delta_{n0}
    +
    \hspace{-0.03cm}
    \int_0^{\tau}\hspace{-0.25cm}\dd t  \dot{\lambda}  \sum_{m} 
    \alpha_m
    \braphi[n]
    (\agp-\exactagp)
    \ketphi[m]
    \hspace{-0.03cm}
    e^
    {i\left(
    \Theta_n-\Theta_m
    \right)}.
\end{equation}

In the scenario where the initial state is the ground state $\ket{\psi(0)}=\ket{\phi_0(0)}$ we are interested in the final ground state fidelity ${\mathcal{F}:=|\bra{\psi(\tau)}\ket{\phi_{0}(\tau)}|^2}$. 
In SM~\cite{supp} we derive the following upper bound for the infidelity 
\begin{equation}
\label{eq:infidelity-general}
    1-\mathcal{F}
    \leq
    4\sum_{n\geq1}
    \int_0^1\dd \lambda  
    \abs{
    \braphi[n]
    (\agp-\exactagp)
    \ketphi[0]
    }^2
    .
\end{equation}
We can additionally use the definitions of the exact and approximate AGPs to rewrite 
$$
    \braphi[n]
    (\agp-\exactagp)
    \ketphi[0]
    =
    i\braphi[n]\dhdl\ketphi[0] 
    [p(\omega_n)-1/\omega_n],
$$
where $\omega_n:=\omega_{n0}$.

We decompose the upper bound of the infidelity shown in Eq.~\eqref{eq:infidelity-general} into two parts, depending on the transition frequency $\omega_n$ (see also Fig.~\ref{fig:poly-fit-one-over-x}). Transitions to states $n\in\indexset$, where $\indexset:= \{n|\,\omega_n\leq\omegamax,\,\forall t\}$, will be addressed by fitting the polynomial $p(\omega_n)$. The residual error $\varepsilon_{\text{res}}$ coming from transitions to states $n\notin\indexset$ will be addressed later.

Let us first focus on the error stemming from transitions to states $n\in\indexset$ for which the polynomial approximation works. Starting from Eq.~\eqref{eq:infidelity-general} we perform a change of variables $t\to\lambda$ and find that this contributes
\begin{equation}
\label{eq:infidelity-poly}
    \varepsilon_{\text{poly}}
    \leq
    4\infnorm{p(\omega)-1/\omega}^2
    \sum_{n\in\indexset}
    \abs{
    \braphi[n]
    \dhdl
    \ketphi[0]}^2.
\end{equation}
We further note that, for the case of $\omegamax=\Omega$, where transitions to all excited states $n\geq 1$ contribute to this part of the error 
$${\sum_{n\geq 1}\abs{
    \braphi[n]
    \dhdl
    \ketphi[0]}^2=\braphi[0](\dhdl)^2\ketphi[0]
    -
    \braphi[0]\dhdl\ketphi[0]^2}
    ,
$$
which is typically extensive for local systems. Furthermore, if we choose $\omegamax=\Omega$ then $\varepsilon_{\text{poly}}$ is the only contribution to the infidelity. This result remains valid for any total protocol time $\tau$ including the limit of $\tau\to 0$, where we only evolve under $\agp$.

However, for $\omegamax < \Omega$ there is a residual error contribution $\varepsilon_{\text{res}}$ coming from transitions $n\notin\indexset$ not addressed by $\agp$. Due to the growth of the constructed polynomial outside of the approximation interval, this term is 
$$
\varepsilon_{\text{res}}
\leq
C
\sum_{n\notin\indexset}
\omega_n^{4d}
\abs{
    \braphi[n]
    \dhdl
    \ketphi[0]}^2,
$$
where we have upper bounded the difference between $p(\omega)$ and $1/\omega$ by $C\omega^{2d}$. This expression ignores potential effects of adiabaticity, which are explicitly accounted for in the SM~\cite{supp}. Moreover, in certain cases the matrix elements might decay rapidly as a function of the transition frequency~\cite{avdoshkin_euclidean_2020, supp}. This might significantly reduce the effects of the residual errors in practice (see also Fig.~\ref{fig:numerics-tau0}).

\section{Krylov construction}

The exact AGP $\exactagp$ for a gapped ground state can be expressed as 
\begin{equation}
    \exactagp 
    =
    \int_0^\infty {\rm d}t g_\Delta(t) e^{-i H t} \dhdl e^{iHt},
\end{equation}
where $g_\Delta(t)$ is some filter function that decays faster than any polynomial and has a scale that is set by the gap~\cite{kolodrubetz_geometry_2017,hastings05,bachmann12,bachmann17}, e.g., following~\cite{bachmann12} we can choose $g_\Delta(t)$ to asymptotically decay almost exponentially like $ \sim \exp{-\Delta t/ \log(t)}$. Next, let us introduce an orthonormal Krylov basis of the Liouvillian $\mathcal{L}=[H,\cdot]$ applied to $\dhdl$. For simplicity we use the inner product $AB = \frac{1}{2^N}{\rm Tr}[A^\dagger B]$. Let us define:
\begin{align}
    O_1 &=\mathcal{L}O_0/b_1 = [H,O_0]/b_1, \nonumber\\
b_1&=\norm{\mathcal{L}O_0},
\end{align}
where $O_0=\partial_\lambda H/ \norm{\partial_\lambda H}$ and proceed for $n>2$ with:
\begin{align}
    O_{n}' & = \mathcal{L}O_{n-1}-b_{n-1}O_{n-2}, \nonumber \\
    O_n & = \frac{O_{n}'}{b_{n}}, \, {\rm with}\, b_n = \norm{O_n'}.
\end{align}
Now we can express the AGP in this Krylov basis as 
\begin{equation}
    \exactagp= \norm{\partial_\lambda H} \sum_{n=0}^\infty O_n \int_0^\infty {\rm d}t g_\Delta(t) \varphi_n(t),
\end{equation}
where $\varphi_n(t)$ are the Krylov expansion coefficients.
As discussed before, all even terms vanish. The approximate AGP, to degree $2d-1$, simply corresponds to the first $2d-1$ terms in the sum. Consequently, the square error $\varepsilon(\agp):=\norm{\exactagp-\agp}^2$ becomes
\begin{equation}
    \varepsilon(\agp)
    =
    \norm{\partial_\lambda H}^2
    \sum_{n=2d}^\infty 
    \left|
    \int_0^\infty {\rm d}t g_\Delta(t) \varphi_n(t)
    \right|^2.
    \label{eq:errorKrylov}
\end{equation}
The convergence is thus determined by the spreading of the operator in Krylov space. The latter has been extensively studied in Ref.~\cite{parker19}. In particular, when $\varphi_n(0)=\delta_{n,0}$ one finds that in generic models the operators spread exponentially fast on the Krylov chain, i.e., $\left<n\right>_t \sim e^{\gamma t}$. However, with $g_\Delta(t)$ decaying (almost) exponentially the error defined by expression~\eqref{eq:errorKrylov} decays with $d$, albeit very slowly. For simplicity, let us apply Cauchy-Schwarz to upper bound Eq.~\eqref{eq:errorKrylov} by
\begin{equation}
    \varepsilon(\agp)\leq
    \norm{\partial_\lambda H}^2 \sum_{n=2d}^\infty \int_0^\infty {\rm d}t g_\Delta(t) \int_0^\infty {\rm d}t g_\Delta(t) |\varphi_n(t)|^2.
\end{equation}
The only timescale in $g_\Delta (t)$ is $1/\Delta$, such that $\int_0^\infty {\rm d}t g_\Delta(t) \sim 1/\Delta$ and using that $|\varphi_n(t)|^2$ is a probability distribution concentrated around $\left<n\right>_t \sim e^{\gamma t}$ one finds 
\begin{equation}
    \int_0^\infty {\rm d}t g_\Delta(t) |\varphi_n(t)|^2 \sim \frac{1}{\gamma} \frac{1}{n^{1+\Delta/\gamma}},
\end{equation}
where we have dropped a $\log\log n$ correction in the exponent coming from the fact that $g_\Delta(t)$ has to decay slower than exponential. As such we find that 
\begin{equation}
    \varepsilon(\agp)
    = \norm{\exactagp-\agp}^2
    \lesssim 
    \frac{\norm{\partial_\lambda H}^2}{\Delta \gamma} (2d)^{-\Delta/\gamma}. 
\end{equation}
Consequently, one requires exponentially large degree in the inverse gap $d\sim e^{c\gamma/\Delta}$ for the error to be small.

\end{appendix}

\iftrue

\pagebreak
\widetext
\begin{center}
\newpage
\textbf{\large Supplemental Materials: Counterdiabatic Driving with Performance Guarantees}
\end{center}

\setcounter{equation}{0}
\setcounter{figure}{0}
\setcounter{table}{0}
\setcounter{section}{0}
\setcounter{page}{1}
\makeatletter
\renewcommand{\theequation}{S\arabic{equation}}
\renewcommand{\thefigure}{S\arabic{figure}}
\renewcommand{\bibnumfmt}[1]{[S#1]}
\section{Additional numerics for the Chebyshev approximation}
Here, we present additional numerical results for the polynomial approximation~\cite{hasson_approximation_2007} presented in the Appendix. In Fig.~\ref{sfig:poly-errs}(a) we show how the error of the constructed polynomials scales as a function of the degree, and verify that it agrees with the analytical prediction. In Fig.~\ref{sfig:poly-errs}(b) we how scaling the gap $\gap$ influences the required degree to stay compliant with a chosen error budget.
\begin{figure*}[htb]
    \centering
    \subfloat[]{
        \includegraphics{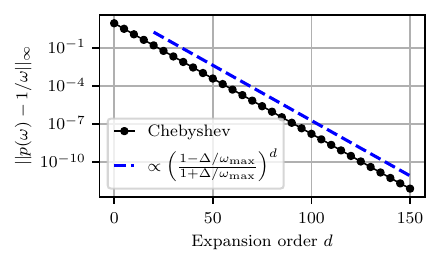}
    }
    \hspace{1cm}
    \subfloat[]{
        \includegraphics{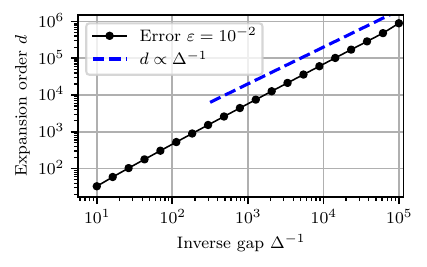}
    }
    \caption{(a) Error of the polynomial approximation for $\Delta=0.1$ and $\omegamax=1$, compared to the analytically predicted scaling. (b) The expansion order $d$ required to construct a polynomial that deviates at most $\varepsilon=10^{-2}$ from $1/\omega$ on the interval $[\Delta, \omegamax]$. For comparison, we plot the asymptotically predicted scaling of the degree.}
    \label{sfig:poly-errs}
\end{figure*}

\section{Deriving the error bound}

Let us derive a bound for the infidelity starting from main text Eq.~\eqref{eq:int-schrodinger}
\begin{equation}
        \alpha_n(\tau)
    =
    \delta_{n0}
    +
    \int_0^{\tau}\dd t  \dot{\lambda}  \sum_{m} 
    \alpha_m
    \braphi[n]
    (\agp-\exactagp)
    \ketphi[m]
    e^
    {i\left(
    \Theta_n-\Theta_m
    \right)}.
    \tag{arXiv-12}
\end{equation}
By an appropriate gauge choice (see Ref.~\cite{chandra_adiabatic_2010}
below Eq.~(4.14)), we can explicitly remove the term with $m=n$ that generates the Berry phase. We are interested in an expression that bounds the infidelity
$
    \mathcal{I}
    =
    \sumgz{n}
    \abs{\alpha_n}^2
$.
For brevity, let us define 
\begin{equation}
    X_{mn}
    =
    \braphi[n]
    (\agp-\exactagp)
    \ketphi[m]
    e^
    {i\left(
    \Theta_n-\Theta_m
    \right)},
\end{equation}
where $X_{mn}=(X_{nm})^*$. We can write
\begin{equation}
    \abs{1-\alpha_0}
    =
    \abs{
    \sumgz{n}
    \int_0^\tau\dd t \dot{\lambda} X_{n0}\alpha_n
    },
\end{equation}
and use $$\abs{1-\alpha_0}\geq 1-\abs{\alpha_0}$$ and 
$$\mathcal{I}=1-\abs{\alpha_0}^2=(1-\abs{\alpha_0})(1+\abs{\alpha_0})\leq 2(1-\abs{\alpha_0})\leq 2\abs{1-\alpha_0}$$
to obtain
\begin{equation}
    \mathcal{I}
    \leq
    2\abs{
    \sumgz{n}
    \int_0^\tau\dd t \dot{\lambda} X_{n0}\alpha_n
    }.
\end{equation}

This bound is valid for any integration time, so in principle also for the time $t_{\max}$ for which $\mathcal{I}$ attains its maximal value $\mathcal{I}_{\max}$. We can perform a change of variables and apply the triangle and Cauchy-Schwartz inequalities to obtain
\begin{equation}
    \mathcal{I}
    \leq
    \mathcal{I}_{\max}
    \leq
    2\sumgz{n}
    \int_0^1\dd\lambda
    \abs{X_{0n}}\abs{\alpha_n}
    \leq
    2\int_0^1\dd\lambda
    \sqrt{\sumgz{n}\abs{X_{n0}}^2}
    \sqrt{\sumgz{n}\abs{\alpha_n}^2}
    \leq
    2\sqrt{\mathcal{I}_{\max}}
    \int_0^1\dd\lambda
    \sqrt{\sumgz{n}\abs{X_{n0}}^2},
\end{equation}
where we have used that the maximal fidelity is upper bounded by the integral over the entire protocol, since the integrand is nonnegative.
Dividing by $\sqrt{\mathcal{I}_{\max}}$ and squaring then yields our main text Eq.~\eqref{eq:infidelity-general}
\begin{equation}
    \mathcal{I}
    \leq
    4\left(
    \int_0^1\dd\lambda
    \sqrt{\sumgz{n}\abs{X_{n0}}^2}
    \right)^2
    \leq
    4\sumgz{n}\int_0^1\dd\lambda\abs{X_{n0}}^2,
\end{equation}
where in the final step we used the Cauchy-Schwarz inequality
$$
\int_0^1 \sqrt{f(\lambda)} \, \dd \lambda 
\leq \left( \int_0^1 f(\lambda) \, \dd \lambda \right)^{\frac{1}{2}} 
     \left( \int_0^1 \dd \lambda \right)^{\frac{1}{2}} 
= \left( \int_0^1 f(\lambda) \, \dd \lambda \right)^{\frac{1}{2}}.
$$

\section{Infidelity at finite time scales}
In this Section we explicitly consider the effects of finite adiabatic timescales. Starting from Eq.~\eqref{eq:infidelity-general} we first decompose the error as discussed in the Appendix, namely by splitting the contributions according to $\indexset$. Following Ref.~\cite{chandra_adiabatic_2010}, the amount of leakage from the ground state into an excited state $k\notin\indexset$ can be approximated as
\begin{equation}
    \abs{\alpha_n}^2
    \approx
    \abs{
    \int_0^{\tau}\dd t  \dot{\lambda}\,
    \alpha_0
    \braphi[n]
    (\agp-\exactagp)
    \ketphi[0]
    e^
    {i\left(
    \Theta_n-\Theta_0
    \right)}}^2.
\end{equation}
In order to obtain analytical expressions we employ methods from adiabatic perturbation theory~\cite{chandra_adiabatic_2010}. We begin by approximating the phase factor $$\Theta_n(t)-\Theta_0(t)=\int_{0}^{t}\dd t' [E_n(t')-E_0(t')]\approx\bar{\omega}_{n}\,t,$$
where we have defined the mean transition frequency $\bar{\omega}_{n}$.
We furthermore make the conservative approximation that $\alpha_0\approx 1$, which results in overestimating the leakage. 
Plugging in the expressions for the matrix elements of the approximate AGP $\agp$ and the exact AGP $\exactagp$ we obtain
\begin{equation}
\label{seq:approximate-pop}
    \abs{
        \alpha_n
    }^2
        \approx
    \abs{
        \int_0^{\tau}\dd t  \dot{\lambda}\,
        \braphi[n]
        \dhdl
        \ketphi[0]
        (p(\omega_n)-1/\omega_n)
        e^
        {
        i\bar{\omega}_nt
        }
    }^2.
\end{equation}
Since we are predominantly interested in the scaling properties of this quantity, we perform the approximation of replacing the matrix element $\braphi[n]\dhdl\ketphi[0]$ with its mean value along the protocol, and replace $\omega_n\to\bar{\omega}_n$. 
This leaves us with
\begin{equation}
     \abs{
        \alpha_n
    }^2
    \approx
    \abs{
    \braphi[n]
    \dhdl
    \ketphi[0]
    }_{\text{avg}}^2
    \,
    \abs{
    p(\bar{\omega}_n)
    -
    1/\bar{\omega}_n
    }^2
    \abs{
    \int_0^{\tau}\dd t  \dot{\lambda}\, e^
        {
        i\bar{\omega}_nt
        }
    }^2,
\end{equation}
where we recognize in $\int_0^{\tau}\dd t  \dot{\lambda}\, e^{i\bar{\omega}_nt}$ the Fourier transform of $\dot{\lambda}$, provided that $\dot{\lambda}$ vanishes $\forall t \notin [0,\tau]$.

\begin{figure*}[htb]
    \centering
    \subfloat[]{
        \includegraphics{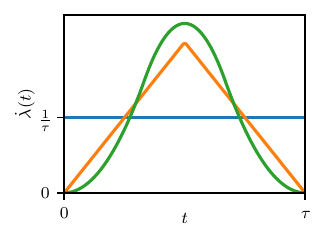}
    }
    \subfloat[]{
        \includegraphics{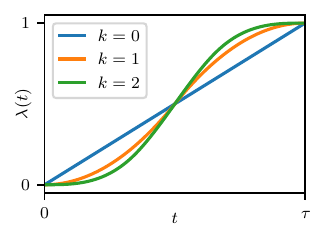}
    }
    \subfloat[]{
        \includegraphics{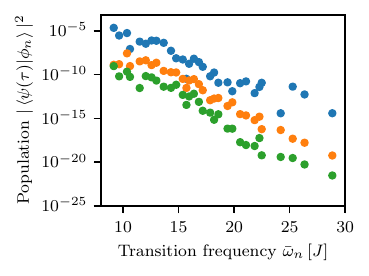}
    }
    \caption{The derivatives (a) and the protocols defined in Eq.~\eqref{seq:protocols} themselves (b), for different convolution orders $k$. In (c) these protocols are used to adiabatically prepare the ground state for the system defined in the main text, for $h_z=J/\varphi$.}
    \label{sfig:lambdas_pops}
\end{figure*}
The expression $\abs{p(\bar{\omega}_n) - 1/\bar{\omega}_n}$ asymptotically scales as $\sim\bar{\omega}_n^{2d}$ at the $d$-th order of the polynomial expansion. Thus, we can compensate for the growth of the polynomial by a suitable choice of $\dot{\lambda}$ that decays sufficiently quickly in frequency space. One possible choice is to consider $k$-fold convolutions of the rectangular window function~\cite{ingham_note_1934} 
$$
W(t, w):=1 \text{ if } 0\leq t\leq w, \text{ else }0,
$$
and setting $w=\tau/k$ to ensure that the length of the protocol is $\tau$. According to the convolution theorem, the Fourier transform of such $k$-fold convolutions is 
\begin{equation}
\mathcal{F}[\underbrace{W(t,\tau/k)*\dots*W(t,\tau/k)}_k]
=
\mathcal{F}[W(t,\tau/k)]^k
=
\left(
\frac{
2\sin{\left(\omega\tau/k\right)}
}{
\omega
}
\right)^k.
\end{equation}
We further need to ensure that $\lambda(\tau)=\int_0^{\tau}\dd t\dot{\lambda}=1$ by proper normalization, obtaining
\begin{equation}
\label{seq:protocols}
\dot{\lambda}(t)
=
\underbrace{W(t,\tau/k)*\dots*W(t,\tau/k)}_k
\cdot
\left(
\frac{k}{\tau}
\right)^k,
\end{equation}
which has Fourier components that scale as $\sim \omega^{-k} (k/\tau)^k$. The protocols are visualized in Fig.~\ref{sfig:lambdas_pops}. Additionally, we provide numerical evidence that increased smoothness of the protocol yields lower eigenstate probabilities for excited states with large $\omega_n$.
The idea is now to choose $k=2d$ to compensate for the growth of the polynomial for $\omega>\omegamax$. However, due to the $(2d/\tau)^{2d}$ factor we furthermore require that $\tau\gtrsim 2d$. 

Ingham~\cite{ingham_note_1934} extended this construction to consider infinitely many convolutions of rectangular window functions $W(t,\tau_k)$, where $\sum_{k=1}^{\infty}\tau_k=\tau$. It can then be shown that the Fourier spectrum of such functions decays as $\sim\exp(-\omega\tau\cdot\xi(\omega))$, where $\int^{\infty}\dd\omega\xi(\omega)/\omega<\infty$. Here, the lower bound of the integral is omitted as we are interested in the convergence of the integral in the $\omega\to\infty$ tail. The function $\xi(\omega)$ can be chosen, e.g., as $\log^2\omega$. Thus, up to logarithmic corrections in the exponent, the spectrum can decay exponentially, where the rate of the decay is controlled by the adiabatic timescale $\tau$. We note, however, that true exponential decay is impossible for functions of compact support. Still, we can leverage this nearly exponential decay when constructing the Laplace polynomials (see below).



\section{Bounds on matrix elements of local operators}
\label{app:matrix-elements}
Here, we will derive an upper bound on the residual error arising from the transition frequencies larger than the upper bound $\omegamax$ of the approximation interval. Concretely, we want to upper bound
$$
\sum_{n\notin\indexset}\hspace{-0.25cm}
\abs{\braphi[n]\dhdl\ketphi[0]}^2
\cdot
\abs{
P_{2d-1}(\omega_{n})
-
\frac{1}{\omega_{n}}
}^2.
$$
We begin by noting that for $\omega_n:=\omega_{n0}>\omegamax$ the difference 
$
\abs{P_{2d-1}(\omega_{n})
-
1/\omega_{n}}
<
C\omega_{n}^{2d-1}
$ 
 for some $C\in\mathbb{R}$.
 
\subsection{Bounding the matrix elements of \texorpdfstring{$\dhdl$}{}}
Next, we use the approach of Avodoshkin \& Dymarsky~\cite{avdoshkin_euclidean_2020} to bound the off-diagonal matrix elements of $\partial_\lambda H$ in the energy eigenbasis. In Ref.~\cite{avdoshkin_euclidean_2020} bounds are placed on the infinity norm of a local operator $O$ by considering its evolution in Euclidian time under a local Hamiltonian $H=\sum_{j=1}^{M} h_j$, where $\infnorm{h_j}\leq J$, namely
\begin{equation}
\label{eq:euclidian-op}
O(-i\beta) 
=
e^{\beta H}
O
e^{-\beta H}
=
\sum_{j=0}^{\infty}
\,
\underbrace{
    [
    H,
    [
    H,\dots
    [
    H,}_{j} O
    ]
    ]
    ]
    \frac{\beta^j}{j!}
\end{equation}
for some real $\beta$. They show that infinity norm of $O(-i\beta)$ is bounded as
\begin{equation}
\label{eq:bound-fbeta}
\infnorm{O(-i\beta)}
\leq
\infnorm{O} f(\beta),
\end{equation}
where $f(\beta)$ is determined in Ref.~\cite{avdoshkin_euclidean_2020} by bounding the norms of commutators in Eq.~\eqref{eq:euclidian-op}.

We can now use their results to place bounds on moments of the ground state power spectrum $\Phi(\omega)$ of $\dhdl$ defined as
$$
\Phi(\omega)
=
\sum_n
\abs{\left(\dhdl\right)_{n0}}^2
    \,\delta\left(\omega_{n} - \omega\right),
$$
where $\left(\dhdl\right)_{n0}=\mel{\phi_n}{\dhdl}{\phi_0}$. In particular, this allows us to write
$$
\sum_{n\notin\indexset}\hspace{-0.25cm}
\abs{\bra{\phi_0(\lambda)}\dhdl\ket{\phi_n(\lambda)}}^2
\left(
P_{2d-1}(\omega_{n})
-
\frac{1}{\omega_{n}}
\right)^2
\leq 
C
\int_{\omegamax}^{\infty}
\dd \omega
\,
\Phi(\omega)
\omega^{2(2d-1)},
$$
where the l.h.s. is nothing but the residual error.
Following the procedure in Section IV of Ref.~\cite{avdoshkin_euclidean_2020} we can further rewrite this as 
$$
\int_{\omegamax}^{\infty}
\dd \omega
\,
\Phi(\omega)
\omega^{2(2d-1)}
=
\sum_{n\notin\indexset} 
\abs{\left(\dhdl\right)_{n0}}^2
\omega_{n}^{2(2d-1)}
=
\sum_{n\notin\indexset}
e^{-2\beta\omega_{n}}
\abs{\left(\dhdl\left(-i\beta\right)\right)_{n0}}^2
\omega_{n}^{2(2d-1)}
.
$$
If we assume that $\omegamax > (2d-1)/\beta$ such that the product of the polynomial and the exponential is largest for the lowest frequency in the sum (i.e., $\omegamax$), then
$$
\sum_{n\notin\indexset}\hspace{-0.25cm}
\abs{\bra{\phi_0(\lambda)}\dhdl\ket{\phi_n(\lambda)}}^2
\left(
P_{2d-1}(\omega_{n})
-
\frac{1}{\omega_{n}}
\right)^2
\leq
C'^2
e^{-2\beta\omegamax}
\omegamax^{2(2d-1)}
\sum_{n\notin\indexset}
\abs{\left(\dhdl\left(-i\beta\right)\right)_{n0}}^2.
$$
We can further bound
$$
\sum_{\omega_{n} \geq \omegamax}
\abs{\left(\dhdl\left(-i\beta\right)\right)_{n0}}^2
\leq
\Tr\left(\ket{\phi_0}\bra{\phi_0}\dhdl(i\beta)\, \dhdl(-i\beta)\right)
\leq
\infnorm{\dhdl(-i\beta)}^2
\leq
\sum_{i=1}^{M}
\infnorm{O_i(-i\beta)}^2,
$$
where, in the final inequality, we used that $\dhdl=\sum_{j=0}^{M}O_j$ is a local operator.
Finally, we use Eq.~\eqref{eq:bound-fbeta} combined with the results of Ref.~\cite{avdoshkin_euclidean_2020} and use the fact that $\forall j\,\infnorm{O_j}\leq K$
\begin{equation}
\label{eq:final_bound_matel}
    \sum_{\substack{n \\ \omega_{n0} > \omegamax}}\hspace{-0.35cm}
\abs{\bra{\phi_0(\lambda)}\dhdl\ket{\phi_n(\lambda)}}^2
\abs{
P_{2d-1}(\omega_{n0})
-
\frac{1}{i\omega_{n0}}
}^2
\leq
\left[
C'
M K \kappa(\omegamax)\omegamax^{2d-1}
\right]^2
\leq
\left[
C'
M K e^{-\gamma\omegamax}\omegamax^{2d-1}
\right]^2
\end{equation}
where we have used that $\kappa$ decays faster than exponentially for one dimensional systems~\cite{avdoshkin_euclidean_2020}. For two-dimensional systems we expect $\kappa\sim \omega\exp(-\gamma'\omega)$ for $\omega\to\infty$.

We note that the derivation of Eq.~\eqref{eq:final_bound_matel} relies on the assumption that $\omegamax > (2d-1)/\beta$, and furthermore the bound in Ref.~\cite{avdoshkin_euclidean_2020} uses $\beta=\log(\omega/4J)/2J$ in the 1D case, meaning that $2d\lesssim\omegamax\log\omegamax$ for this derivation to be valid.

\begin{figure*}[htb]
    \centering
    \subfloat[]{
        \includegraphics{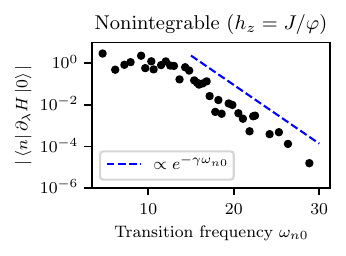}
        \label{sfig:exp-small-matels}
    }
    \subfloat[]{
        \includegraphics{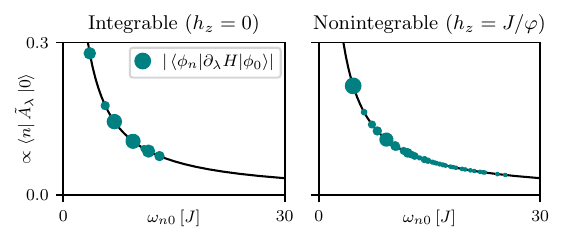}
        \label{sfig:matels}
    }
    \caption{(a) The average magnitude of the matrix elements of $\dhdl$ for the nonintegrable Ising chain from the main text. (b) The distribution of the transitions along the protocol, highlighting the qualitiative differences in the spectra of the integrable and nonintegrable cases. The marker size is set according to the average strength of the coupling $\abs{\braphi[n]\dhdl\ketphi[0]}$ along with the protocol, displayed as a function of the mean transition frequency along the protocol. $J$ has been set to unity.}
    \label{sfig:int-vs-nonint-structure}
\end{figure*}

\section{Variational Counterdiabatic Driving}

Variational counterdiabatic driving~\cite{sels_minimizing_2017} provided a variational method to determine the coefficients of local expansions of the exact adiabatic gauge potential. Given an ansatz $\agp=\sum_k c_k O_k$ a cost function $S[\agp]=\Tr(G[\agp]^2)$ is proposed where $G[\agp]=\dhdl-i[H,\agp]$. It can be shown that the function $S[\agp]$ attains its minimum precisely at the exact AGP $\agp=\exactagp$, providing a method to obtain the coefficients $c_k$.

If the nested commutator ansatz is chosen for $\agp$ (cf. Eq.~\eqref{eq:nested_commutator_expansion}) we can gain deeper insights into the nature of the variational approach~\cite{claeys_floquet-engineering_2019}. We begin by defining the response function 
$$
    \Gamma(\omega, X)
    =
    \sum_{n,m}
    \abs{\braphi[n]X\ketphi[m]}^2
    \delta(\omega - \omega_{mn}),
$$
and its moments as
$$
    \Gamma^{(k)}(X)
    =
    \int_0^{\infty}\dd\omega\,
    \Gamma(\omega, X)\omega^{2k}.
$$
Then it can be shown that minimizing the quadratic objective $\Tr[G^2]$ can be equivalently represented as a system of linear equations~\cite{claeys_floquet-engineering_2019}
$$
\sum_{k=1}^d c_k \Gamma^{(k + j)}(\dhdl)
=
\Gamma^{(j)}(\dhdl),
\quad
\text{for }
j=1\dots d.
$$
Solving this linear system of equations can be used to obtain the coefficients without the need for variational optimization~\cite{xie_variational_2022}. However, the system quickly becomes ill-conditioned with increasing $d$, requiring careful resolution by e.g., a modified Gram-Schmidt process (see also Appendix on the Krylov construction).

Additionally, it has been shown~\cite{claeys_floquet-engineering_2019} that the variationally obtained operator $\agp$ at the $d$-th order matches the first $d$ moments of the response exact AGP's response function
$$
    \Gamma(\omega,\agp)^{(k)}
    =
    \Gamma(\omega,\exactagp)^{(k)}, 
    \quad\text{for }
    k=1\dots d.
$$

\section{Laplace polynomials}

As indicated in the main text, an alternative approach is to define a family of polynomials that is orthogonal with respect to the weight function $w(\omega)=e^{-\gamma\abs{\omega}}$, which is the unnormalized Laplace distribution. Therefore we dub the polynomials \emph{Laplace polynomials} and denote them via the symbol $\Lambda_k$. These polynomials should satisfy the orthogonality relation
$$
\langle
\Lambda_k,
\Lambda_j
\rangle_w
:=
\int_{-\infty}^{\infty} \dd\omega\,
\Lambda_k(\omega)
\Lambda_j(\omega)
w(\omega)
=
\delta_{kj}.
$$
Since these polynomials have not been extensively studied analytically, we here construct them numerically using the \texttt{Wolfram Mathematica} package \texttt{OrthogonalPolynomials}~\cite{cvetkovic2004mathematica, milovanovic2012special}. We leverage the fact that 
$$
\int_{-\infty}^{\infty} \dd\omega\,
\omega^n w(\omega)
=
(1 + (-1)^n)\, 
n!
/
\gamma^{n+1}
$$
to numerically construct the three term recurrence relation for the polynomials, which facilitates their evaluation. 

\begin{figure*}[htb]
    \centering
    \subfloat[]{
        \includegraphics{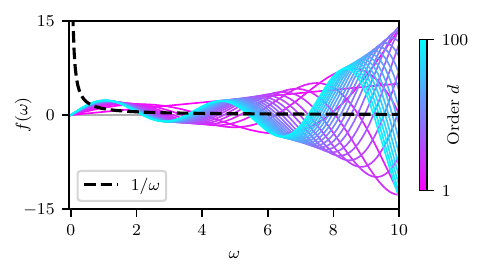}
        \label{sfig:laplace_polynomials}
    }
    \subfloat[]{
        \includegraphics{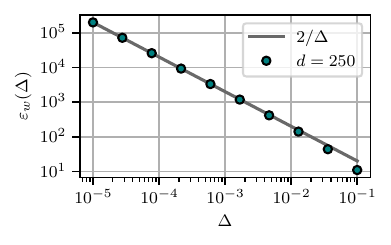}
        \label{sfig:laplace_gap_error}
    }
    \caption{The odd Laplace polynomials fit to the function $1/\omega$ are shown in (a) for $\gamma=1$. In (b) the error $\varepsilon_w$ is shown at $d=250$ as a function of the gap $\gap$. The fit $2/\gap$ nicely describes the observed data.}
    \label{sfig:laplace-1}
\end{figure*}

We next fit the polynomials $\Lambda_{2k-1}(\omega)$ to the function $1/\omega$ by computing the expansion coefficients as ${c_{2k-1}=\langle\Lambda_{2k-1},1/\omega\rangle_w}$. We plot the resulting polynomials $p(\omega)=\sum_{k=1}^d c_{2k-1}\Lambda_{2k-1}(\omega)$ in Fig.~\ref{sfig:laplace_polynomials}. We are now interested in how the error of the approximation with respect to the weight function $w$ behaves as a function of the gap, i.e., we are interested in the scaling properties of the quantity
\begin{equation}
\label{seq:epsilon_w}
    \varepsilon_w (\gap)
    =
    2\int_{\Delta}^{\infty}\dd\omega\,
    \left(
        1/\omega - p(\omega)
    \right)^2
    w(\omega),
\end{equation}
which we display in Fig.~\ref{sfig:laplace_gap_error}, and where the factor $2$ stems from the even parity of the integrand and the symmetric domain. In principle, typically the weight function $w$ will only kick-in after some (constant) local energy scale $J$; however, we expect this not to affect the asymptotic findings presented here.

We find that asymptotically for small $\gap$ the error $\varepsilon_w$ is well described by $2/\gap$, which originates from the integral of the $w(\omega)/\omega^2$ term in Eq.~\eqref{seq:epsilon_w}. Note that the asymptotic scaling of this term is independent of $\gamma$. Consequently, it is sensible to rescale the errors as $\varepsilon_w \cdot\gap/2$.

\begin{figure*}[htb]
    \centering
    \subfloat[]{
        \includegraphics{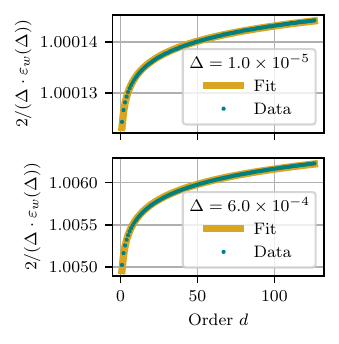}
        \label{sfig:lap_errors_fit}
    }
    \subfloat[]{
        \includegraphics{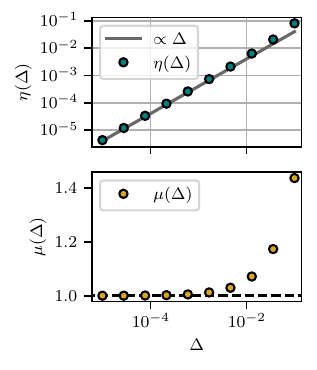}
        \label{sfig:etamu}
    }
    \caption{In (a) the reciprocal and rescaled error is shown together with the fit Eq.~\eqref{seq:fit}. In (b) the scaling of fitted coefficients $\eta$ and $\mu$ with $\gap$ is displayed. All plots have been generated for $\gamma=1$.}
    \label{sfig:laplace-2}
\end{figure*}

The analytical results of the Krylov construction presented in the Appendix motivate the functional dependence 
\begin{equation}
    \label{seq:fit}
    f(d)=\mu(\Delta)e^{\eta(\Delta) \log d}.
\end{equation}
Indeed, we see in Fig.~\ref{sfig:lap_errors_fit} that the reciprocal rescaled error is well-described by this fit. We find that $\eta(\gap)\propto\gamma\gap$ and that $\mu(\gap)\to 1$ as $\gap\to 0$ (see Fig.~\ref{sfig:etamu}).

Combining these insights we find that asymptotically the error scales as
\begin{equation}
    \varepsilon_w
    \sim
    (2/\gap)\cdot e^{-c\gamma\Delta\log d} 
\end{equation}
which, for $d\gg 1$, means that the expansion order $d$ has to scale as $d\sim\exp\left(\frac{\log(1/\Delta)}{\gamma\Delta}\right)$. Thus, we are able to reproduce the results from the Krylov construction (see Appendix) up to logarithmic factors in the exponent.
\fi

\end{document}